\documentstyle[preprint,eqsecnum,aps,epsfig]{revtex}

\begin{document}
\draft
\title{Description of even-even triaxial nuclei within 
the Coherent State and the Triaxial Rotation Vibration Models}
\author{U. Meyer$^a$, 
A.A. Raduta$^{a,b}$ and 
 Amand Faessler$^a$}
\address{$^{a)}$Institut f\"{u}r Theoretische Physik der Universit\"{a}t
T\"{u}bingen, 
Auf der Morgenstelle 14, D-72076 T\"{u}bingen,  Germany}
\address{$^{b)}$Institute of Physics and Nuclear Engineering, 
Bucharest, POBox MG6, Romania}
\date{\today}
\maketitle
\begin{abstract}
The coherent state model (CSM) and the triaxial rotation-vibration 
model (TRVM) are alternatively used to describe the ground, 
$\gamma$ and $\beta$ bands 
of $^{228}$Th. CSM is also applied to the nuclei $^{126}$Xe and
$^{130}$Ba, which were recently considered in TRVM. 
The two models are compared with respect to both their underlying assumptions
and to their predicted results for energy levels and E2 branching ratios. 
Both models describe energies and quadrupole transitions of $^{228}$Th 
equally well and in good agreement with experiment,
if the 0$_3^+$ level at 1120 keV is interpreted as the 
head of the $\beta$ band. The other two 0$^+$ levels at 832 and 939 keV are
most likely not of a pure quadrupole vibration nature as has been
already pointed out in the literature.
\end{abstract}
\narrowtext
\section{Introduction}
\label{sec:level1}
The quadrupole degrees of freedom have been intensively used by 
phenomenological models to interpret the data for
energies and electromagnetic
transitions of collective states. In the pioneering model of Bohr and Mottelson~\cite{bohr}
 ({\bf L}iquid {\bf D}rop {\bf M}odel=LDM)
some collective
properties are treated in terms of quadrupole shape coordinates describing
small oscillations of the nuclear surface around a spherical equilibrium
shape.\par
The harmonic motion of the liquid drop and the restriction to a spherical
shape for the ground state are severe limitations of this approach.
 The first improvement of the LDM was obtained in the rotation-vibration
model (RVM)~\cite{fae1,fae2,fae3,fae4} in which the deviation of the 
shape coordinates from
their static values is considered and by this an axially symmetric
deformed shape is described. Anharmonicities were introduced by Greiner
and Gneuss~\cite{greiner}. In this way many collective features for
the complex spectra could be explained consistently.\par
In order to explain quantitatively the excitation energies and transition
probabilities, the interacting boson approximation (IBA) exploits 
underlying group symmetries \cite{ArIach,ArIach1}. In IBA most of the nuclei have been ordered
in three categories characterised by the dynamical symmetries of the
model Hamiltonian. The symmetries correspond to the groups O(6), SU(5) 
and SU(3) and the specific features are $\gamma$-instability, $\gamma$-stability
and the quasi degeneracy of states with equal angular momentum belonging to
$\gamma$ and $\beta$ bands, respectively.\par
The coherent state model (CSM), developed in the beginning of the eighties,
 treats an effective Hamiltonian in a restricted model space generated by
projecting out the angular momentum from three orthogonal and deformed 
states~\cite{radu81,radu82,radu831,radu183,radu87,Ann,Roum,Roum1,radu96,radu97}. 
 These are chosen as the lowest elementary boson excitations of
an axially symmetric coherent state. The axially deformed picture is
very convenient since it allows to define the $K$ quantum number.\par
Recently, there appeared several data about triaxial deformed nuclei which
were interpreted in the IBA by using its O(6) limit~\cite{Kir}. This was 
a real challenge for the RVM authors who extended the model to 
triaxial equilibrium shapes ({\bf T}riaxial {\bf R}otation-{\bf V}ibration
{\bf M}odel=TRVM)~\cite{ulinpa}. The TRVM was applied to the nuclei
$^{126}$Xe and $^{130}$Ba and the results obtained were in equally good
agreement with available data \cite{Kir,Siems,seiffert} as 
in IBA~\cite{ulinpa} .\par
A short while ago, new data for $^{228}$Th were 
detected~\cite{gar98}. According to these data, $^{228}$Th behaves as
a nucleus without axial symmetry. Since the first successful applications
of CSM referred to the Pt region \cite{radu81}, which corresponds to the O(6) symmetry 
in the IBA interpretation, it is expected that triaxial nuclei like
$^{126}$Xe, $^{130}$Ba and $^{228}$Th  can be realistically  described by CSM.
We furthermore apply the TRVM for the new data of $^{228}$Th.\par
We are not only interested  in the predictions of the two models but we
would also like to point out resemblances and differences of their theoretical
ingredients. Therefore we discuss in Section 4, after a brief description 
of the two approaches in Sections 2 and 3, a possible relation between the
two models. We present our numerical results in Section 5. Section 6 contains
the final conclusions.
\section{The Triaxial Rotation Vibration Model}
\label{sec:level2}
The RVM was extended recently~\cite{ulinpa} to triaxial nuclei (TRVM).
 Here we briefly present the results.\par
The TRVM describes small oscillations of a quadrupole deformed nuclear 
surface
\begin{equation}
R(\theta,\phi)=R_0(1+\sum_{\mu}\alpha_{2\mu}Y_{2\mu}^*(\theta,\phi)),
\end{equation}
around an ellipsoidal shape without axial symmetry. The classical 
Hamiltonian governing the motion of the quadrupole shape 
coordinates $\alpha_{2\mu}$ is given by:
\begin{equation}\label{Ham}
H=\frac{1}{2}B\sum_{\mu}\dot{\alpha}_{2\mu}^*\dot{\alpha}_{2\mu}+V(\alpha_{2\mu})\hspace*{0.2cm}.
\end{equation}
In the intrinsic reference frame the five degrees of freedom are $a_0$,
 $a_2$ and $\Omega$, where $\Omega$ denotes the Euler angles fixing the
position of the intrinsic frame with respect to the laboratory frame.
 The $a_k$ ($k$=0,2) are obtained from the coordinates $\alpha_{2k}$ 
through the rotation $\hat{R}(\Omega)$:
\begin{equation}
a_k=\hat{R}(\Omega)\alpha_{2k}\hat{R}(\Omega)^{\dag}\hspace*{0.3cm},
\hspace*{0.3cm}k=0,2\hspace*{0.2cm}.
\end{equation}
In the Bohr-Mottelson parametrization, the new coordinates $a_k$ 
are expressed in terms of $\beta$ and $\gamma$ deformations by:
\begin{eqnarray}
a_0(t) & = & \beta \mbox{cos}(\gamma),\nonumber\\
a_2(t) & = & \frac{1}{\sqrt{2}}\beta \mbox{sin}(\gamma).
\end{eqnarray}
The stationary points of the trajectory defined by ($a_0(t)$, 
$a_2(t)$, $\Omega(t)$) are identical with stationary points of the
equipotential energy surface. Assuming ($\beta_0$, $a_2$) to be a minimum
of this surface , one may deduct from the surface variables
$a_0$ and $a_2$ their static parts and consider resulting deviations
as new dynamical variables:
\begin{eqnarray}
a_0(t) & = & \beta_0+a_0^{\prime}(t),\\
a_2(t) & = & a_2+a_2^{\prime}(t).
\end{eqnarray}
The new dynamical coordinates are supposed to be small comparing them to the
static deformations. In this case we may expand the model
Hamiltonian up the second order in $a_0^{\prime}/\beta_0$ and 
$a_2^{\prime}/a_2$.\par
After quantization, the Hamiltonian splits up into several terms
which can be written as:
\begin{equation}\label{Ham1}
H\equiv T+V=T_{rot}+T_{vib}+T_{rotvib}+V_{\beta_0 a_2}(a_0^{\prime}, 
a_2^{\prime}),
\end{equation}
where the following notations have been used:
\begin{eqnarray}\label{Ham2}
T_{rot} & = & \frac{\hat{{\bf I}}^2-\hat{I}_3^2}{2I_0}+
\frac{\hat{I}_3^2}{16Ba_2^2},~~
T_{vib} = -\frac{\hbar^2}{2B}(\frac{\partial^2}{\partial a_0^{\prime 2}}+
\frac{1}{2}\frac{\partial^2}{\partial a_2^{\prime 2}}),\nonumber\\
T_{rotvib} & = &   \frac{\hat{{\bf I}}^2-\hat{I_3}^2}{2I_0}f_0(\beta_0,a_2,a_0^
{\prime},a_2^{\prime})
+ \frac{\hat{I}_{+}^2+\hat{I}_{-}^2}{2I_0}f_1(\beta_0,a_2,a_0^
{\prime},a_2^{\prime})\nonumber\\
{} & {} &
+ \frac{\hat{I}_3^2}{16Ba_2^2}f_2(a_2,a_2^{\prime})+2\epsilon
\frac{a_0^{\prime}}{\beta_0},
\\ \nonumber
V(a_0^{\prime},a_2^{\prime}) & = & \frac{1}{2}C_0a_0^{\prime^2}+
C_2a_2^{\prime^2}.
\end{eqnarray}
The final expressions for the coefficients $f_0$, $f_1$, and $f_2$, 
obtained by the above mentioned expansion, are :
\begin{eqnarray}
f_0 & = & -2\frac{a_0^{\prime}}{\beta_0}+3\frac{a_0^{\prime^{2}}}{\beta_0^2}+
\frac{2}{\beta_0^2}(a_2^2+2a_2a_2^{\prime}+a_2^{\prime^2}),
\nonumber\\
f_1 & = & \frac{1}{3}\sqrt{6}\frac{1}{\beta_0}(a_2+a_2^{\prime})-
\sqrt{6}\frac{1}
{\beta_0^2}a_0^{\prime}(a_2+a_2^{\prime}),\nonumber
\\
f_2 & = & -2\frac{a_2^{\prime}}{a_2}+3\frac{a_2^{\prime^2}}{a_2^2}.
\end{eqnarray}
The moment of inertia ($I_0$) is equal to $I_0=3B\beta_0^2$ and its inverse
is denoted by $\epsilon=1/I_0$.\par
The eigenstates
of the unperturbed Hamiltonian
\begin{equation}
H_0=T_{rot}+T_{vib}+V,
\end{equation}
are taken as diagonalization basis for the coupling Hamiltonian.
A basis state $\vert IK,n_2n_0\rangle$ is labelled by the total angular
momentum ($I$), its projection on the intrinsic $z$-axis ($K$) and by the
number of phonons for the $\beta$ ($n_0$) and $\gamma$ ($n_2$) vibrations.
 For $K$=0, the angular momentum $I$ takes only even values, whereas for
$K$=2,4,6,$\ldots$ all values $I>K$ are allowed. The basis is restricted
to quantum numbers $K\leq 6$ and $n_2+n_0\leq 2$.\par
The TRVM has four parameters. These are the vibration energies 
$E_{\beta}(=\hbar \sqrt{\frac {C_0}{B}})$
and $E_{\gamma}(=\hbar \sqrt{\frac {C_2}{B}})$, the inverse moment of inertia $\epsilon$ and the
ratio $a_2/\beta_0$ of the static deformations. To these four
parameters, we add the Lipas's parameter $\alpha_L$~\cite{lipas}, which 
corrects the 
incomplete description of the variation of the moment of inertia
due to the restriction of the diagonalization space.
 The Lipas's parameter relates the excitation energies $E_0$,
obtained by diagonalizing  the model Hamiltonian,  with  
the energies $E$ which are to be compared with the data:
\begin{equation}
E=E_0/(1+\alpha_L E_0).\hspace*{1cm}
\end{equation}
The Lipas's parameter influences only the energies, but not the
wavefunctions.\par
The transition probabilities can be readily obtained once we have determined 
the initial and final states
as well as the transition operator. In Ref.~\cite{ulinpa} a compact 
expression for the transition operator $m(E2,\mu)$, was obtained. This is
given by:
\begin{eqnarray}
m(E2,\mu)& = &\frac{3Z}{4\pi}R_0^2\left[{\cal D}_{\mu 0}^2\left(\beta_0
\left(1+\frac{2}{7}\left(\frac{5}{\pi}\right)^{\frac{1}{2}}\beta_0\right)
\right)+{\cal D}_{\mu 0}^2 a_0^{\prime}\left(1+\frac{4}{7}
\left(\frac{5}{\pi}\right)^{\frac{1}{2}}\beta_0\right)+\right.\nonumber\\
{}&{}& {\cal D}_{\mu 0}^2\frac{2}{7}\left(\frac{5}{\pi}\right)^{\frac{1}{2}}
({a_0^{\prime}}^2-2(a_2+a_2^{\prime})^2)+\\
{}&{}&\left.
({\cal D}_{\mu 2}^2+
{\cal D}_{\mu -2}^2)\left(\left(1-\frac{4}{7}\left(\frac{5}
{\pi}\right)^{\frac{1}{2}}\beta_0\right)(a_2+a_2^{\prime})-
\frac{4}{7}\left(\frac{5}{\pi}\right)^{\frac{1}{2}}a_0^{\prime}(a_2+
a_2^{\prime})\right)\right]\nonumber
\end{eqnarray}
We use standard notations for the nucleus charge ($Ze$), nuclear 
radius ($R_0$) and Wigner's functions (${\cal D}^2_{MK}$).
 The transition operator depends on both the static and
the dynamical deformations. It contains not only terms which are linear  
in $a_0^{\prime}$ and $a_2^{\prime}$ but also quadratic and constant terms.
 While the latter terms are caused by the deformation effects due to the
transformations (2.5) and (2.6), the former
terms reflect an anharmonic structure for the E2 transition.
\section{The Coherent State Model}
\label{secòlevel 3}
The coherent state model (CSM) was to a 
great part developed, in collaboration, by one of the present
authors (A.A.R.)\cite{radu82}, with the scope to describe the main features of the
collective ground, $\beta$ and $\gamma$ bands.\par
First, one defines a restricted collective space by projecting out the
components with good angular momentum from three orthogonal deformed states.
 One of the states, $\Psi_g$, is a coherent state with axial symmetry 
describing a deformed ground state. The remaining states, denoted  
by $\Psi_{\gamma}$ and $\Psi_{\beta}$, are obtained by exciting $\Psi_g$
with polynomials of second and third rank in the quadrupole bosons. These
are chosen in such a way that the three deformed states are mutually 
orthogonal. Moreover, one requires that the orthogonality is preserved
after projection was performed.\par
The three states depend on a real parameter ($d$) which simulates the
nuclear quadrupole deformation. Indeed, $d$ is proportional to the
expectation value of the quadrupole moment operator corresponding to the 
deformed ground state. As a matter of fact due to this property the
attribute 'deformed state' may be assigned to the three states before 
projection.\par
In the vibrational limit ($d\to$0) \cite{Roum,Roum1}, the states projected from 
$\Psi_g$, $\Psi_{\gamma}$ and $\Psi_{\beta}$ are the highest seniority
states, whereas in the rotational limit $(d\geq3)$ \cite{Ann} they behave 
similarly 
as the liquid drop wavefunctions for the ground state, $\gamma$ and $\beta$ 
bands, respectively. The intermediate situations where $K$ is not a good quantum number
are reached by a smooth variation of the deformation parameter $d$. 
 In this way one achieves a one to one correspondence between vibrational and
rotational states which agrees with the semi-empirical rule
of Sheline and Sakai \cite{Shel,Sak}. \par
In the restricted quadrupole boson space spanned by the projected states 
one determines an effective Hamiltonian which ideally should be diagonal in
the model basis states. The simplest solution is a sixth order boson
Hamiltonian which has vanishing off diagonal matrix elements for the $\beta$ 
band states:
\begin{equation}\label{csmhamilton}
H=A_1\left(22\hat{N}+5\Omega_{\beta^{\prime}}^{\dag}\Omega_{\beta^{\prime}}
\right)+A_2\hat{I}^2+A_3\Omega_{\beta}^{\dag}\Omega_{\beta}.
\end{equation}
Here $\hat{N}$ and $\hat{I}^2$ denote the quadrupole boson number
and total angular momentum squared operators, respectively. 
The other notations are:
\begin{eqnarray}
\Omega_{\beta^{\prime}}^{\dag} & = & \left[b_2^{\dag}\times b_2^{\dag}
\right]^0-\frac{d^2}{\sqrt{5}},\\
\Omega_{\beta}^{\dag} & = & \left[b_2^{\dag}\times b_2^{\dag}\times b_2^{\dag}
\right]^0+\frac{3d}{\sqrt{14}}\left[b_2^{\dag}\times b_2^{\dag}\right]^0
-\frac{d^3}{\sqrt{70}},
\end{eqnarray}
where $b_{2\mu}^{\dag}$ ($-2\leq\mu\leq$2) are the components of the quadrupole
boson operator. Consequently, the energies of the $\beta$ and $\gamma$ 
band states of odd angular momentum are given as expectation values 
on the corresponding projected states. The energies of ground state
and gamma band states with even angular momenta are obtained by
diagonalizing a 2$\times$2 matrix. We would like to emphasize that
the off-diagonal matrix elements vanish in the extreme regimes.\par
The reduced E2 transition probabilities are described by the anharmonic
operator:
\begin{equation}
Q_{2\mu}=q_1\left(b_{2\mu}^{\dag}+(-1)^{\mu}b_{2-\mu}\right)+q_2
\left(\left[b_2^{\dag}b_2^{\dag}\right]^2_{\mu}+
\left[b_2b_2\right]^2_{\mu}\right).
\end{equation}
Rotational and vibrational limits for energies and B(E2) values
have been studied analytically in Refs.\cite{Ann,Roum,Roum1}\par
Numerical calculations showed that the CSM describe equally well nuclei 
of different symmetries like O(6) ($^{190,192,194}$Pt), SU(3) ($^{232}$Th,
 $^{238}$U) and SU(5) ($^{150}$Sm, $^{152}$Gd). In some 
cases, for example for $^{156}$Dy, the $\beta$ band has a complex structure
which can be described after adding two more terms to the CSM-Hamiltonian:
\begin{equation}
H^{\prime}=A_4\left(\Omega_{\beta}^{\dag}\Omega_{\beta^{\prime}}+h.c.\right)
+A_5\Omega_{\beta^{\prime}}^{\dag\hspace*{0.1cm}2} 
\Omega_{\beta^{\prime}}^{2}.
\end{equation}
which do not alter the decoupling property of the $\beta$ band.
 The CSM was extended by including the coupling of quadrupole collective
motion to the individual \cite{radu831,radu183,radu87} as well as to the 
octupole degrees of freedom \cite{Ann,radu96,radu97}.
CSM differs from the IBA formalism in several essential features. Indeed, the
model Hamiltonian is a boson number non-conserving Hamiltonian. The CSM
states are projected from an infinite series of bosons and therefore 
dynamical effects for any nuclear deformation and angular momentum can be 
accounted for. In particular, due to the coherent property of the deformed
ground state, the CSM works very well for high spin states. We would like
to note that the collective motion within CSM is determined by
both high anharmonicities involved in the model Hamiltonian and the
complex structure of the model space. The CSM is compared with the TRVM 
in the next Section.
\section{Comparison between CSM and TRVM}
\label{sec:level4}
Since the deformed ground state is a vacuum state for the shifted 
quadrupole boson operator
\begin{equation}
(b_{20}-d)\Psi_g=0,
\end{equation}
and moreover the transformation $e^T$ with 
$T=d\left(b_{20}^{\dag}-b_{20}\right)$ produces such a shift 
\begin{equation}
e^Tbe^{-T}=b-d\hspace*{0.2cm},\hspace*{0.2cm}
e^Tb^{\dag}e^{-T}=b^{\dag}-d,
\end{equation}
one expects that the transformed Hamiltonian $e^THe^{-T}$ is a deformed 
operator which may describe the motion around an axially
deformed shape.

The classical motion of an axially non-symmetric shape can be studied
by the associated classical energy function:
\begin{equation}
{\cal H}=\langle\Psi\vert\hat{H}\vert\Psi\rangle,
\end{equation}
where 
\begin{equation}
\vert\Psi\rangle=exp(z_0b_{20}^{\dag}+z_2b_{22}^{\dag}+z_{-2}b_{2-2}^{\dag}
-z_0^{\star}b_{20}-z_2^{\star}b_{22}-z_{-2}^{\star}b_{2-2})\vert 0\rangle,
\end{equation}
and $\hat{H}$ is given by (\ref{csmhamilton}). The vacuum state for the
quadrupole bosons is denoted by $\vert 0\rangle$. The coefficients
$z_0, z_2, z_{-2}, z_0^{\star}, z_2^{\star}, z_{-2}^{\star}$ are complex
functions of time and define the classical phase space 
coordinates.
 By direct calculation one finds:
\begin{eqnarray}
{\cal H} & = & 2\left(11A_1+3A_2\right)\left(z_0z_0^{\star}+z_2z_2^{\star}
+z_{-2}z_{-2}^{\star}\right)+
\nonumber\\{}&{}&A_1\left(z_0^2+2z_2z_{-2}-d^2\right)
\left(z_0^{\star\hspace*{0.1cm}2}+2z_2^{\star}z_{-2}^{\star}-d^2\right)+
\nonumber\\
{} & {} & \frac{A_3}{70}\left[2\left(6z_0z_2z_{-2}-z_0^3\right)
+3d\left(z_0^2+2z_2z_{-2}\right)-d^3\right]
\nonumber\\{}&{}&\left[2\left(
6z_0^{\star}z_2^{\star}z_{-2}^{\star}-z_0^{\star\hspace*{0.1cm}3}
\right)+3d\left(z_0^{\star\hspace*{0.1cm}2}
+2z_2^{\star}z_{-2}^{\star}\right)-d^3\right].
\end{eqnarray}
The motion of the phase space coordinates is governed by the equations 
provided by the variational principle\footnote{We use the units 
$\hbar$=c=1}
\begin{equation}
\delta\int\langle\Psi\vert(H-i\frac{\partial}{\partial t}\vert\Psi\rangle
dt^{\prime}=0.
\end{equation}
The results are:
\begin{equation}
\left\{z_k,{\cal H}\right\}=\dot{z}_k\hspace*{0.2cm},
\hspace*{0.2cm}\left\{z_k^{\star},{\cal H}\right\}=\dot{z}^{\star}_k
\hspace*{0.2cm},
\hspace*{0.2cm}\left\{z_k,z_{k^{\prime}}^{\star}\right\}=
-i\delta_{kk^{\prime}},
\end{equation}
where the Poisson bracket is defined with respect to the canonically
conjugate variables ($\sqrt{2}Re(z_k), \sqrt{2}Im(z_k)$).\par
Stationary points of this motion  are also stationary points
for  the surface of constant energy:
\begin{equation}
{\cal H}(z_0, z_0^{\star}, z_2, z_2^{\star}, z_{-2}, z_{-2}^{\star})=E
\end{equation}
Suppose now that this surface exhibits a minimum point $(z_0, z_2)=
(u_0, u_2)$ with $u_0$ and $u_2$  being real numbers. The existence of such a 
minimum is proved in Ref.~\cite{radu98}. Since we want to mention
here some classical features which do not depend on whether this minimum 
is axially symmetric or not we consider the simplifying case $u_2$=0.
 Expanding ${\cal H}$ around the minimum point and keeping only the
quadratic terms in the deviations $z_k^{\prime}, z_k^{\star\hspace*{0.1cm}
\prime}$, one obtains:
\begin{equation}
{\cal H}={\cal H}_0+{\cal H}_{\beta}+{\cal H}_{\gamma}.
\end{equation}
where ${\cal H}_0$ is a constant term (not depending on coordinates) and
\begin{eqnarray}
{\cal H}_{\beta} & = & B_1z_0^{\prime}z_0^{\star\hspace*{0.1cm}\prime}+
B_2\left(z_0^{\prime\hspace*{0.1cm}2},
+\left(z_0^{\star\hspace*{0.1cm}\prime}\right)^2\right)~,\\
{\cal H}_{\gamma} & = & G_1\left(z_2^{\prime}z_2^{\star\hspace*{0.1cm}\prime}
+z_{-2}^{\prime}z_{-2}^{\star\hspace*{0.1cm}\prime}\right)+G_2
\left(z_2^{\prime}z_{-2}^{\prime}+
z_2^{\star\hspace*{0.1cm}\prime}z_{-2}^{\star\hspace*{0.1cm}\prime}\right),
\end{eqnarray}
The coefficients $B_1, B_2, G_1$ and $G_2$ are explicitly given in 
Appendix~A. Note that at the level of quadratic approximation
there is no $\beta-\gamma$ coupling term. \par
The classical motion can be quantized as follows. One defines first
a new set of coordinates and momenta:
\begin{eqnarray}
Q_2=\frac{1}{\sqrt{2}}(z_2^{\star}+z_{-2}), & 
Q_{-2}=\frac{1}{\sqrt{2}}(z_{-2}^{\star}+z_{2}), & 
Q_0=\frac{1}{\sqrt{2}}(z_0^{\star}+z_{0}),\\
P_2=\frac{i}{\sqrt{2}}(z_{-2}^{\star}+z_{2}), & 
P_{-2}=\frac{i}{\sqrt{2}}(z_{2}^{\star}+z_{-2}), & 
P_0=-i\frac{\partial}{\partial Q_0}.
\end{eqnarray}
The Poisson brackets of these coordinates can be easily calculated and the 
result reflects
their canonically conjugate character:
\begin{equation}
\left\{Q_k,P_{k^{\prime}}\right\}=\delta_{kk^{\prime}}\hspace*{0.2cm},
\hspace*{0.2cm}k=0,\pm 2\hspace*{0.2cm}.
\end{equation}
For small deviations from the minimum point, it is useful to introduce the 
parametrization:
\begin{equation}
Q_{\pm 2}=\frac{\gamma}{\sqrt{2}}\hspace*{0.1cm}e^{\pm 2i\phi}
\equiv \frac{q_{\pm 2}}{\sqrt{2}}
\end{equation}
Let us consider as coordinate operators the $Q_k$ as defined above, but 
denoted hereafter by $\hat{Q}_k$,
and the corresponding momenta defined by:
\begin{equation}
\hat{P}_{\pm 2} = -\frac{i}{\sqrt{2}}\frac{\partial}{\partial q_{\pm 2}}
\hspace*{0.2cm},\hspace*{0.2cm}\hat{P}_0=-i\frac{\partial}{\partial Q_0}
\end{equation}
Indeed, one can easily check that
\begin{equation}
\left[\hat{Q}_{\pm 2},\hat{P}_{\pm 2}\right]=1
\end{equation}
The transformation ($Q_{\pm 2},P_{\pm 2}$)$\rightarrow$($\hat{Q}_{\pm 2},
\hat{P}_{\pm 2}$) is usually called canonical quantization. Quantized
Hamiltonians are obtained by writing ${\cal H}_{\beta}$ and ${\cal H}_{\gamma}$
in terms of ($Q_{\pm 2}, P_{\pm 2}$) and then making the 
above mentioned replacements. The latter transformation is made after 
putting the mixed $Q$ and $P$ terms in a symmetrized form. The final
results are:
\begin{eqnarray}
\hat{H}_{\gamma} & = & -\frac{1}{2}(G_1-G_2)(\frac{\partial^2}
{\partial\gamma^2}+\frac{1}{\gamma}\frac{\partial}{\partial\gamma}+\frac{1}
{4\gamma^2}\frac{\partial^2}
{\partial\phi^2})+\frac{G_1+G_2}{2}\gamma^2\\
\hat{H}_{\beta} & = & -\frac{1}{2}\left(B_1-2B_2\right)
\frac{\partial^2}{\partial Q_0^2}+\frac{1}{2}(B_1+2B_2)Q_0^2
\end{eqnarray}
The spectra which are obtained with the above two 
operators, $H_{\beta}$ and $H_{\gamma}$, are:
\begin{eqnarray}
E_{N}^{(\gamma)} & = & \omega_{\gamma}(N+1)\hspace*{0.2cm},\hspace*{0.2cm}
N=2n+\frac{1}{2}\vert K\vert\hspace*{0.2cm},\hspace*{0.2cm}n=0,1,2,\ldots
\hspace*{0.2cm}\vert K\vert=0,2,4,\ldots\\
E_n^{(\beta)} & = & \omega_{\beta}\left(n+\frac{1}{2}\right)\hspace*{0.2cm}
,\hspace*{0.2cm}n=0,1,2,\ldots
\end{eqnarray}
where 
\begin{equation}
\omega_{\gamma}=\left(G_1^2-G_2^2\right)^{1/2}\hspace*{0.2cm},\hspace*{0.2cm}
\omega_{\beta}=\left(B_1^2-4B_2^2\right)^{1/2}
\end{equation}
Inspecting the expressions from Appendix A, one sees that $G_1$ and $B_1$  
are mainly given by the $\hat{N}$ term of the
model Hamiltonian. Indeed, the coefficient $A_2$ accompanying 
$\hat{I}^2$ is usually small. Therefore, the $\gamma$ and $\beta$ harmonic
frequencies are decreased by anharmonicities.\par
Although we discuss the simplest case, the $\gamma$ degree of freedom 
could not be decoupled entirely from the rotational coordinate
$\phi$. This seems to be a general feature for the liquid drop model.\par
The Hamiltonian ${\hat H}_{\beta}+{\hat H}_{\gamma}$ is similar to the 
part $T_{vib}+T_{rotvib}+V$ of the TRVM except for the coupling terms. Indeed, here
only one coupling term is reproduced. The reason is 
that by restricting 
the coherent state to the $b_{20}^{\dag}, b_{22}^{\dag}, b_{2-2}^{\dag}$
bosons, the motion of two Euler angles is not taken into account. If we
considered the coherent states of five variables as the variational state,
the TRVM Hamiltonian would have been entirely obtained, as the harmonic limit, 
by the procedure described above.\par
The fact that the classical picture leads to the  
interpretation of the generating deformed states
given below is remarkable. Under circumstances of small 
deviations from the equilibrium shape the one beta and one gamma phonon states
can be written in the form of $\Psi_{\beta}$ and $\Psi_{\gamma}$, 
respectively \cite{radu98}. Thus, one may assert that by means of projection 
technique the CSM builds up rotational bands on the top of three deformed states 
which represent the ground, beta and gamma one phonon states, respectively. 
The CSM uses a highly anharmonic Hamiltonian and the projected states are
superpositions of all $K$ quantum numbers. This results in generating
some important effects for high spin states. By contradistinction, TRVM uses
a harmonic vibrational Hamiltonian and a diagonalization basis
subject to the restriction  $K\leq 6$. On the other hand, one can easily
describe multi-phonon states as well as $K\neq 0,2 $  rotational bands 
in the TRVM.
To adapt the CSM to higher excited bands would require a great amount of 
additional effort.
\section{Numerical Results}
\label{sec:level5}
The CSM and TRVM have been applied to three triaxial nuclei:
$^{126}$Xe, $^{130}$Ba and $^{228}$Th. Although the TRVM has been 
already considered for the first two nuclei, for the sake of a 
complete comparison between the two models we also invoke these results
in the present paper. 

The parameters obtained through a fitting procedure have been 
collected in Table 1. The TRVM fixes its parameters so that a best overall 
fit for the experimental excitation energies is obtained. The CSM fitting 
procedure is as follows. For a given $d$, $A_1$ and $A_2$ are determined so 
that the excitation energies for $2^+_g$ and $2^+_{\gamma}$ are equal to 
the corresponding experimental data.  
The parameter $A_3$ is used to fit the 
excitation energy for $0^+_{\beta}$. 
Finally, one keeps that value of $d$ which 
assures an overall best fit for the excitation energies in the three bands.
The CSM uses a transition operator which 
depends on two parameters, $q_0$ and $q_2$. Since one deals with branching 
ratios, only one is needed. Therefore we give here the ratio 
$\frac{q_2}{q_0}$.
The branching ratios depend also on the deformation parameter $d$, which was 
already determined from energy analyses.

Note that for $^{126}$Xe and $^{130}$Ba, the best fit is obtained, 
within TRVM,
if one assumes that the $0^+_2$ band is a $K^\pi=0^+$ gamma band .
Another feature revealed by the TRVM parameters consists of that, 
except for $^{130}$Ba,
where $E_\beta$ and $E_\gamma$ are nearly degenerate, the beta band is higher
in energy than gamma band, which reflects a gamma unstable picture or,
in other words  saying, an $O(6)$ symmetry. The property mentioned 
above for the excepting case recommends 
$^{130}$Ba as a representative of $SU(3)$. As seen from the row corresponding 
to the parameter $d$, $^{126}$Xe and $^{130}$Ba are weakly deformed nuclei while 
$^{228}$Th is a well deformed nucleus. The static values of the $\gamma$ deformation,
extracted from the $a_2/\beta _0$ values given in the Table 1, are about 
$25^0$ for
$^{126}$Xe and $^{130}$Ba and $13^0$ for $^{228}$Th.

Although CSM uses a $b_{20}^+$-coherent state for the 
unprojected ground state, 
the $I$-projected states are superpositions of the
functions ${\cal D}^{I^{\star}}_{MK}$ with the expansion coefficients 
$A_{\alpha IK}(\beta \gamma)$ depending on the $\beta$ and $ \gamma$ deformations.
Integrating over $\beta $, the square of these coefficients, 
a probability distribution for the gamma 
variable is obtained. Such investigations can also be performed in 
connection with the projected 
states associated to beta and gamma bands. Since the details of this 
analysis can be found in Ref. \cite{radu832}, we only enumerate the main 
results.
The  $0^+_g$ state behaves as a gamma unstable wavefunction while the ground 
band states of high spin, like $10^+_g$, has a gamma probability with a maximum 
at $45^0$. The $\gamma$ band state 2$_{\gamma}^+$ describes
 a triaxial nucleus ($\gamma=30^0$)
while increasing the spin the probability distribution in the variable 
$\gamma$ goes gradually to the situation when it
has two equal maxima at $0^0$ and $60^0$. In conclusion, the $\gamma$ 
asymmetric features are in CSM 
accounted for  due to the mixture of ground and gamma projected states 
as well as due to the specific anharmonic structure of the model Hamiltonian.

It is instructive to compare the experimental results for the sum of energies 
for the first two states of angular momentum two and the energy of the first
$3^+$ state. The differences are about 2,~47 and 95 keV for $^{228}$Th, 
$^{126}$Xe
and $^{130}$Ba , respectively. These figures suggest that $^{228}$Th could be
described as an asymmetric rotator \cite{dav} and therefore a decoupling 
regime for the $\beta$ band is expected.
The deviation from the rotor picture for $^{126}$Xe and $^{130}$Ba reflects
an additional interaction between the states of ground and gamma bands.

Predicted and experimental energies of ground and gamma bands 
are plotted in Figures 1-3. Since only few data are 
available for the beta band, we summarise the results in Table 2. 
As seen from Table 2, within the CSM the energy spacings in the beta band are
too 
large and this happens due to the magnitude of the $A_2$ coefficient. In 
this case the inclusion
of one additional term from $H^\prime$ (3.5) is necessary. Indeed considering
only the $A_4$ term and fixing its strength as to fit the energy of the $4^+$
state the final results for the other states in the beta band 
are close to those given by the TRVM.
In Ref. \cite {gar98},
 three $K^\pi=0^+$ bands at 832, 939 and 1120 keV have been 
identified. In both models, TRVM and CSM, 
the available data
for transition probabilities could be fairly well described when 
the $0^+_3$ band is interpreted as the beta band. This is consistent with 
some earlier 
investigations \cite{dal87} pointing to the fact that the two lower bands, 
with $K^\pi=0^+$, 
are mainly two octupole phonon and two quasiparticle states. Indeed, 
taking into account that the first $3^-$ state lies 
at 396 keV, the state $0^+_1$ has an excitation energy close to that 
characterising the two octupole phonon state.

It is worth noting that by an language abuse the bands considered here are 
called  $K^\pi$ bands. Indeed, in both models, K is not a good quantum number, 
the eigenstates being superpositions of several K-components.
However in this superposition one K prevails and furthermore this K is taken 
as a band label.

An interesting feature, related to the gamma band, concerns the staggering 
of the $(I^+,(I+1)^+)$ states with $I\geq3$ in $^{126}$Xe and $^{130}$Ba. 
This 
appears in the low part of the spectrum  and more pronounced in $^{130}$Ba.
In CSM the staggering is a reminiscence of the vibrational limit 
(see Fig.3 of Ref. 
\cite{radu82}) where the staggered states are degenerate. Beyond a critical 
value of $d$, the staggering $(3^+,4^+),(5^+,6^+),(7^+,8^+)$,.., etc. which 
characterises the near vibrational states, changes to $(2^+,3^+),(4^+,5^+),
(6^+,7^+)$,..., etc. which is specific for the rotational limit.
In TRVM the staggering is caused by the rotation vibration coupling terms.
The spacings of the lowest two doublets are larger than those shown by 
experiment, in both models. We could decrease these spacings, in CSM,  
choosing a smaller deformation 
parameter $d$ with the price of perturbing some branching ratios. However the 
agreement with experiment is better in the high spin region while for TRVM
the discrepancies increase with angular momentum.

Within TRVM and CSM, the branching ratios characterising the decay of states 
belonging to the aside bands $\beta$ and $\gamma$ are described by means of 
the transition operators (2.11) and (3.4) respectively, with the parameters 
determined from energy analyses. In the case of CSM, there is a parameter 
more, $\frac{q_2}{q_0}$, which is fixed so that the experimental data for the 
branching ratio $(2^+_\gamma \rightarrow0^+_g)/(2^+_\gamma \rightarrow2^+_g)$
is reproduced.

Note that the CSM does not include the anharmonic term
$[b^{\dagger} b]^2_\mu$ in the expression of the transition operator. 
The reason is
that this term gives vanishing contribution  to transitions between beta and 
ground bands. Also the transitions characterising the gamma band are affected 
only by a negligible amount.
The results of our calculations and experimental data are given, 
for comparison, 
in Tables 3-5. The notations we used in these Tables are the standard ones.
Thus for $^{126}$Xe and $^{130}$Ba, $I^+_1, I^+_2$, with I-even, stand for 
states of ground and gamma bands respectively, while $I^+_3$ (I-even) is a 
state from the $K^\pi=0^+$-gamma band in the TRVM and from beta band in 
the CSM.
If $I$ is odd, $I^+_1$ is a state from the gamma band. As we have already 
mentioned ,
$^{228}$Th exhibits three excited $K^\pi=0^+$ bands. The best agreement for
branching ratios are obtained interpreting the $0^+_3$ band as the beta band.
This implies the following notations for $^{228}$Th.
The states $I^+_4$ with $I$=even belong to the gamma band while the states of 
beta 
band are of the type $I^+_5$. Exception is for the state $2^+$ of the gamma 
band which is the third state of angular momentum 2.

From tables 3-5 one may see that both models describe reasonably well the 
data reffering to the gamma band. The agreement quality for the
two models are comparable. One should mention that within TRVM,
branching ratios associated to the state
$2^+_3$ of $^{126}$Xe and $^{130}$Ba are much smaller than the 
corresponding data, 
while CSM predictions lie quite close to the experimental data.
In $^{130}$Ba, the normalised transitions 
$4^+_3\rightarrow2^+_2,~4^+_3\rightarrow2^+_1$ predicted by TRVM exceed the 
experimental data 
by a factor of about 231 and 132 respectively while CSM results 
underestimate the data by a factor of 5 and 4.3, respectively.
These data suggest that additional terms in the TRVM Hamiltonian are 
necessary in order to improve the structure of the eigenfunctions describing
the above mentioned decaying states.
The results obtained in both models for $^{228}$Th agree quite well with 
each other as well 
as with the experimental data. Moreover, the realistic description of the 
data concerning 
the decay of the states $2^+_5$ and $4^+_5$ support our assignment of the 
beta band to the experimental $0^+_3$ band. 
\section{Conclusions}
\label{sec:level6}
In the previous sections two phenomenological models, CSM and TRVM, have been
successively applied to three triaxial nuclei : $^{126}$Xe, $^{130}$Ba 
and $^{228}$Th. While the second model was adapted for triaxial nuclei in a 
previous publication, the original CSM was applied without any modification.
One suggests a possible relation between the two models. Indeed, the TRVM
seems to be the classical counterpart of the CSM in the harmonic limit.
The proof for this relationship has the virtue of suggesting a way of
supplementing the Hamiltonian characterising the TRVM with some anharmonic 
terms. 

The two models yield similarly good results concerning the excitation energies
and transition probabilities for $^{228}$Th. Moreover they are at par 
concerning the interpretation of the $0^+_3$ band as the beta band.
For the remaining two nuclei the considered models are at 
variance with respect to the interpretation of the first 
excited 0$^+$ band. Indeed 
the TRVM describes the first excited $K^\pi=0^+$ band as a gamma band,
i.e. build upon a gamma vibration state, while within the CSM
this is a beta band. This difference in interpretation has an echo in the 
predicted branching ratio characterising the states of this band. One suggests
that including anharmonicities in the TRVM Hamiltonian, the discrepancies
for branching ratios might be removed. 

While TRVM can be easily used for describing higher K-bands build on the top 
of a many phonon state, the extension of the CSM to several bands
requires a good deal of additional work.

\section{Appendix A}
\label{sec:level7}
Here we give the explicit expressions of the coefficients $G_1, G_2, B_1, B_2$
involved in the equations (4.18), (4.19) defining the $\gamma$ and $\beta$
vibrations.
\begin{eqnarray}
G_1&=&22A_1+6A_2, \nonumber \\
G_2&=&2A_1(u_0^2-d^2)-\frac {3A_3}{35}(d+2u_0)(u_0-d)(2u_0-du_0-d^2),
\nonumber \\
B_1&=&22A_1+6A_2+4A_1u_0^2+\frac {18}{35}A_3u_0^2(d-u_0)^2, \nonumber \\
B_2&=&A_1(u_0^2-d^2)-\frac {3A_3}{70}(d-2u_0)(u_0-d)(2u_0-du_0-d^2).
\end{eqnarray}

\vfill\eject
{\bf {\centerline {Table Captions}}}
\vskip0.5cm
{\bf Table 1.}
Parameters used by the TRVM (the first 5) and the CSM (the last 5)
to describe the data for $^{126}$Xe, $^{130}$Ba and $^{228}$Th.
Their significance is explained in the text.
\vskip0.3cm
{\bf Table 2.}
Experimental (first column) and predicted excitation energies 
(in units of keV) of the 0$_2^+$ band given
by the TRVM (second column) and the CSM (third column). 
\vskip0.3cm
{\bf Table 3.}
B(E2) branching ratios for $^{126}$Xe in the triaxial Rot-Vib Model (TRVM)
(first column) and in the Coherent State Model (CSM) (third column) compared 
with experiment (second column)
\vskip0.3cm
{\bf Table 4.}
The same as in Table 3 but for $^{130}$Ba.
\vskip0.3cm
{\bf Table 5.}
The same as in Table 3 but for $^{228}$Th.
\vfill\eject
{\bf {\centerline {Figure Captions}}}
\vskip0.5cm
{\bf Fig. 1.} The experimental (exp.), the TRVM and CSM predicted 
excitation 

energies for $^{126}$Xe are 
represented in units of keV for ground ({\bf a)})  and gamma bands ({\bf b)}).
\vskip0.3cm
{\bf Fig. 2.} The same as in Fig. 1 but for $^{130}$Ba.
\vskip0.3cm
{\bf Fig. 3.} The same as in Fig. 1 but for $^{228}$Th.

\vfill\eject
\begin{center}
{\bf Table 1}\\[24pt]
\renewcommand{\arraystretch}{1.2}
\begin{tabular}{|l||r|r|r|}
\hline
{} & $^{126}$Xe &  $^{130}$Ba &  $^{228}$Th \\
\hline
\hline
$E_{\beta}$ (keV) & 1769 & 1168 & 1120 \\
$E_{\gamma}$ (keV) & 1314 & 1179 & 645 \\
$\epsilon$ (keV) & 81.3 & 80.5 & 17.54\\
$a_{20}/\beta_0$ & 0.333 & 0.329 & 0.1585\\
$\alpha_{L}$ (keV$^{-1}$) & 10$^{-4}$ & 10$^{-4}$ & 10$^{-4}$\\
$d$ & 1.46 & 1.32 & 3.14 \\
$A_{1}$ (keV) & 14.325 & 15.783 & 17.731\\
$A_{2}$ (keV) & 19.211 & 12.377 & 1.512\\
$A_{3}$ (keV) & 14.411 & -0.423 & -7.021\\
$q_2/q_0$ & -0.119 & 0.073 & -0.071\\
\hline
\end{tabular}
\end{center}
\vfill\eject
\begin{center}
{\bf Table 2}\\[24pt]
\renewcommand{\arraystretch}{1.2}
\begin{tabular}{||r||c|c|c||c|c|c||c|c|c||}
\hline
{} & \multicolumn{3}{c||}{$^{126}$Xe}&\multicolumn{3}{c||}{$^{130}$Ba}
& \multicolumn{3}{c||}{$^{228}$Th}\\
\cline{2-10}
{} & Exp. & TRVM & CSM & Exp. & TRVM & CSM & Exp. & TRVM & CSM\\
\hline
0$^+$& 1314 & 1314 & 1314 & 1179 & 1179 & 1179 & 1120 & 1120 & 1120\\
2$^+$& 1679 & 1600 & 1670 & 1557 & 1490 & 1523 & 1176 & 1173 & 1170\\
4$^+$& 2042 & 2150 & 2254 &      & 2053 & 2017 & 1290 & 1292 & 1283\\
6$^+$&      & 2758 & 3025 &      & 2644 & 2635 &      & 1468 & 1454\\
8$^+$&      & 2917 & 3968 &      & 3196 & 3365 &      & 1690 & 1679\\
10$^+$&     & 3518 & 5074 &      & 3947 & 4200 &      & 1946 & 1949\\
\hline
\end{tabular}
\end{center}
\vfill\eject
\begin{center}
{\bf Table 3}\\[24pt]
\begin{tabular}{|r|r|r|r|}
\hline
$I_{i}\rightarrow I_{f}$&TRVM
&{}exp.~\cite{seiffert}&CSM\\
\hline
$2_{2}^{+}\rightarrow 0_{1}^{+}$&8.7&$1.5\pm 0.4$&1.5\\
$2_{1}^{+}$&100.0&100.0&100.0\\
$3_{1}^{+}\rightarrow 2_{2}^{+}$&100.0&100.0&100.0\\
$4_{1}^{+}$&24&$34.0^{+10}_{-34}$&13\\
$2_{1}^{+}$&4.4&$2.0^{+0.6}_{-1.7}$&1.25\\
$4_{2}^{+}\rightarrow 2_{2}^{+}$&100.0&100.0&100.0\\
$4_{1}^{+}$&66&$76.0\pm 22.0$&76\\
$2_{1}^{+}$&1.5&$0.4\pm 0.1$&5.5\\
$0_{2}^{+}\rightarrow 2_{2}^{+}$&100.0&100.0&100.0\\
$2_{1}^{+}$&1.1&$7.7\pm 2.2$&1.1\\
$2_{3}^{+}\rightarrow 0_{2}^{+}$&100.0&100.0&100.0\\
$2_{2}^{+}$&0.8&$2.2\pm 1.0^{*}$&3.14\\
$4_{1}^{+}$&0.04&$2.0\pm 0.8$&0.21\\
$2_{1}^{+}$&0.01&$0.14\pm 0.06^{*}$&0.18\\
$0_{1}^{+}$&0.01&$0.13\pm 0.04$&0.04\\
$3_{1}^{+}$&20&$67.0\pm 22.0^{*}$&20\\
$5_{1}^{+}\rightarrow 6_{1}^{+}$&43&$75\pm 23$&25\\
$4_{2}^{+}$&83&$76\pm 21$&84\\
$3_{1}^{+}$&100.0&100.0&100.0\\
$4_{1}^{+}$&0.8&$2.9 \pm 0.8$&0.15\\
$6_{2}^{+}\rightarrow 6_{1}^{+}$&22&$34^{+15}_{-25}$&64\\
$4_{2}^{+}$&100.0&100.0&100.0\\
$4_{1}^{+}$&1.3&$0.49\pm 0.15$&9.99\\
$7_{1}^{+}\rightarrow 6_{2}^{+}$&22&$40\pm 26$&31\\
$5_{1}^{+}$&100.0&100.0&100.0\\
\hline
\end{tabular}
\end{center}
\vfill\eject
\begin{center}
{\bf Table 4}\\[24pt]
\begin{tabular}{|r|r|r|r|}
\hline
{}&{}&{}&\\
$I_{i}\rightarrow I_{f}$&TRVM&exp.~\cite{Kir}&CSM\\
{}&{}&{}&\\
\hline
$2_{2}^{+}\rightarrow 0_{1}^{+}$&4.5&$6.2\pm 0.7$&6.2\\
$2_{1}^{+}$&100.0&100.0&100.0\\
\hline
$3_{1}^{+}\rightarrow 2_{2}^{+}$&100.0&100.0&100.0\\
$4_{1}^{+}$&16&$22.0\pm 3.0$&23\\
$2_{1}^{+}$&2.6&$4.5\pm 0.6$&9.0\\
\hline
$4_{2}^{+}\rightarrow 2_{2}^{+}$&100.0&100.0&100.0\\
$4_{1}^{+}$&57&$54.0\pm 10.0$&181\\
$2_{1}^{+}$&2.75&$2.3\pm 0.4$&0.29\\
\hline
$0_{2}^{+}\rightarrow 2_{2}^{+}$&100.0&100.0&100.0\\
$2_{1}^{+}$&3.0&$3.3\pm 0.2$&2.2\\
\hline
$2_{3}^{+}\rightarrow 0_{2}^{+}$&100.0&100.0&100.0\\
$2_{2}^{+}$&0.04&$21.0\pm 4.0$&20.13\\
$4_{1}^{+}$&0.11&$2.7\pm 0.5$&1.49\\
$2_{1}^{+}$&0.06&$3.3\pm 0.6$&0.005\\
$0_{1}^{+}$&0.04&$0.017\pm 0.003$&0.016\\
$3_{1}^{+}$&18&${}$&59\\
\hline
$4_{3}^{+}\rightarrow 2_{2}^{+}$&670&$2.9(5)$&0.56\\
$3_{1}^{+}$&214&$97(17)$&17\\
$4_{2}^{+}$&100.0&$100.0^{*}$&100.0\\
$4_{1}^{+}$&7.2&$3.4(6)^{*}$&0.47\\
$2_{1}^{+}$&23.3&$0.30(6)$&0.07\\
\hline
\end{tabular}
\end{center}
\vfill\eject
\begin{center}
{\bf Table 5}\\[24pt]
\begin{tabular}{|r|r|r|r|}
\hline
{}&{}&{}&\\
$I_{i}\rightarrow I_{f}$&TRVM
&{}exp.~\cite{gar98}&{}{}{}{}CSM\\
{}&{}&{}&\\
\hline
$2_{3}^{+}\rightarrow 0_{1}^{+}$ & 54 & 45(3)& 45\\
$2_{1}^{+}$&100.0&100.0&100.0\\
$4_{1}^{+}$&6.1&3.1(3)&6.4\\
\hline
$3_{1}^{+}\rightarrow 2_{1}^{+}$ & 100.0 & 100.0 & 100.0\\
$4_{1}^{+}$&62&67(6)&80\\
\hline
$4_{4}^{+}\rightarrow 2_{1}^{+}$ & 15.1 & 15.0(1.2)& 3.25\\
$4_{1}^{+}$&100.0&100.0&100.0\\
$6_{1}^{+}$&8.84 & 6.2(1.4) & 1.73\\
\hline
$5_{1}^{+}\rightarrow 4_{1}^{+}$ & 100.0 & 100.0 & 100.0\\
$6_{1}^{+}$& 112 & 142(32)& 169\\
\hline
$2_{5}^{+}\rightarrow 0_{1}^{+}$ & 65 & 41(10)& 59\\
$2_{1}^{+}$&100.0&100.0&100.0\\
$4_{1}^{+}$&267&420(60)&187\\
\hline
$4_{5}^{+}\rightarrow 4_{1}^{+}$ & 100.0 & 100.0 & 100.0\\
$6_{1}^{+}$&362&470(240)&149\\
\hline
\end{tabular}
\end{center}
\vfill\eject
\center
{\bf Figure 1a)}\\[24pt]
\epsfig{file=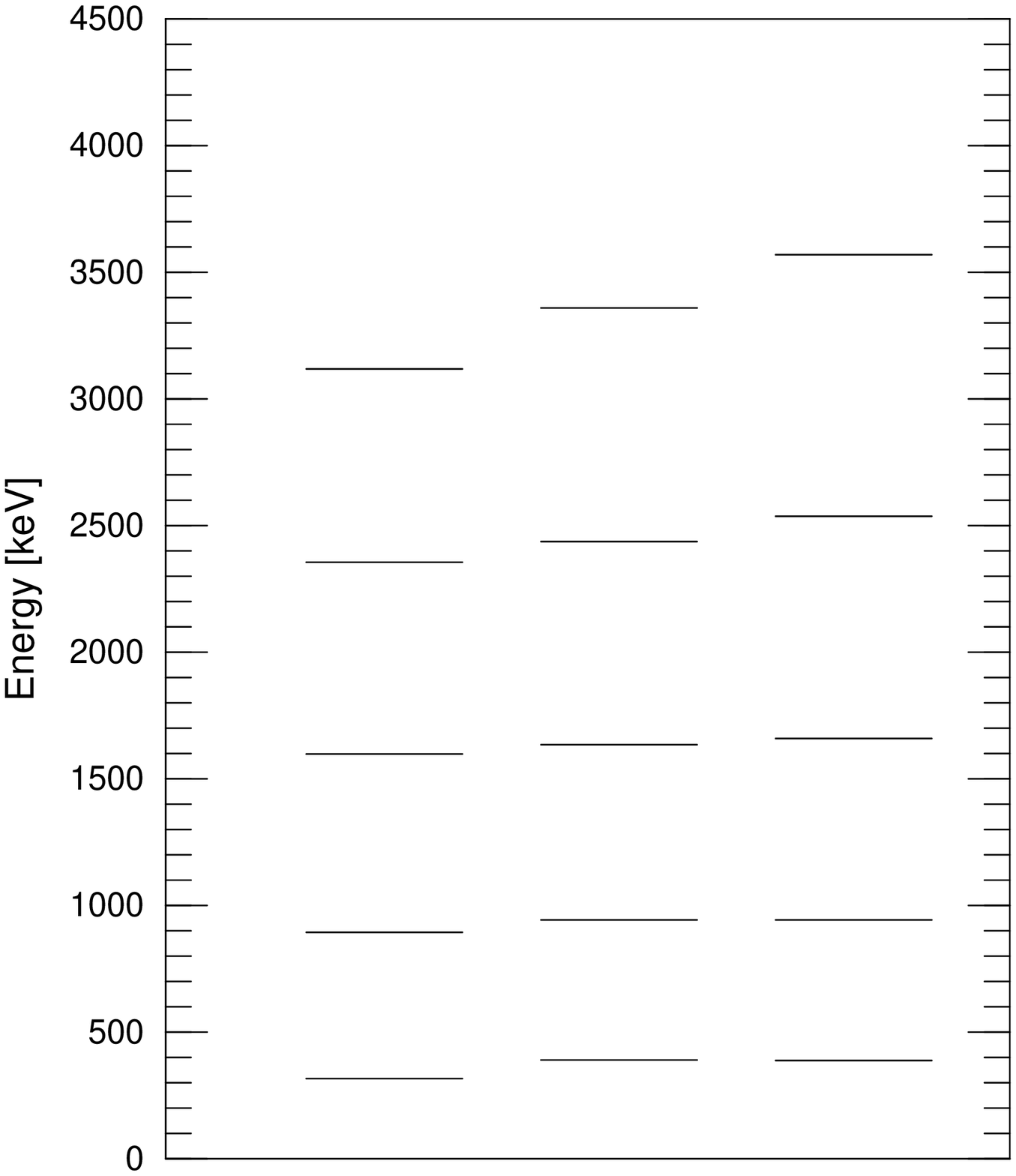,width=0.60\textwidth}
\put(-233,26){2$^{+}$}
\put(-233,70){4$^{+}$}
\put(-233,125){6$^{+}$}
\put(-233,180){8$^{+}$}
\put(-233,235){10$^{+}$}
\put(-208,285){{\bf TRVM}}
\put(-135,285){\bf exp.\cite{Kir}}
\put(-175,315){\large{\bf $^{126}$Xe g.s. band}}
\put(-65,285){{\bf CSM}}
\put(-225,315){{\large a)}}
\vfill\eject
\center
{\bf Figure 1b)}\\[24pt]
\epsfig{file=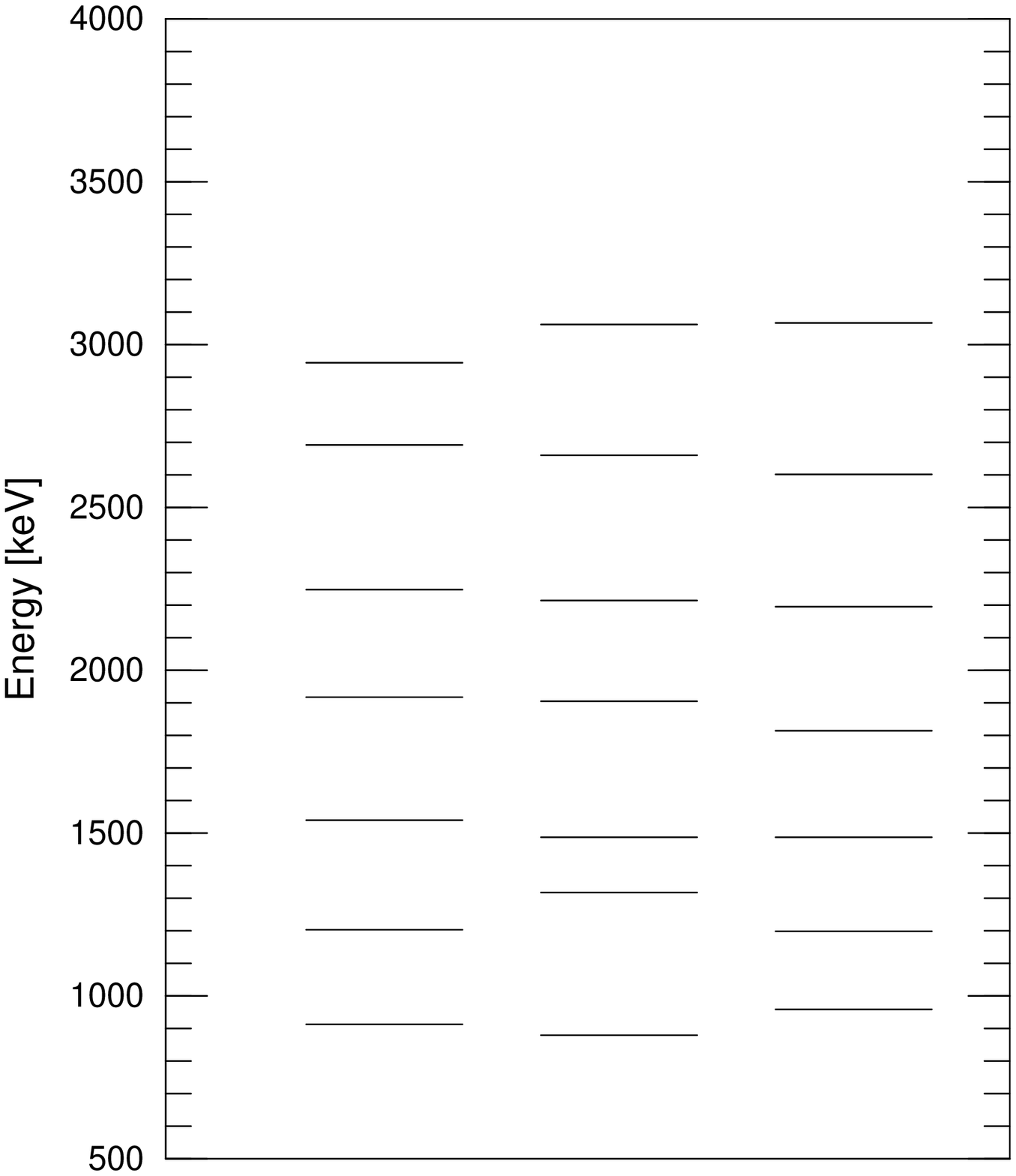,width=0.60\textwidth}
\put(-233,41){2$^{+}$}
\put(-233,70){3$^{+}$}
\put(-233,102){4$^{+}$}
\put(-233,140){5$^{+}$}
\put(-233,170){6$^{+}$}
\put(-233,210){7$^{+}$}
\put(-233,245){8$^{+}$}
\put(-208,280){{\bf TRVM}}
\put(-135,280){\bf exp.\cite{Kir}}
\put(-190,315){\large{\bf $^{126}$Xe $K$=2 $\gamma$ band}}
\put(-65,280){{\bf CSM}}
\put(-225,315){{\large b)}}
\vfill\eject
\center
{\bf Figure 2a)}\\[24pt]
\epsfig{file=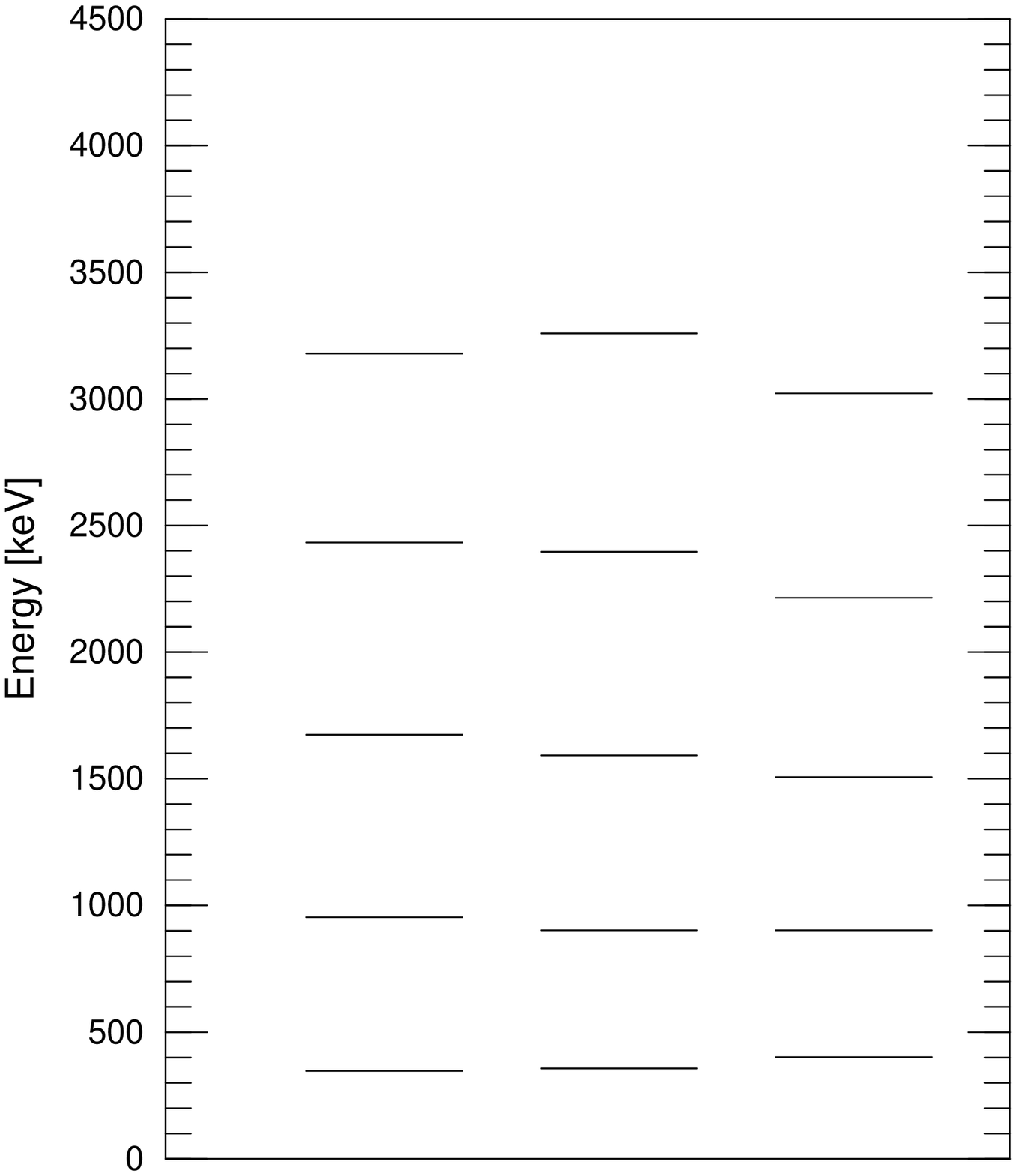,width=0.60\textwidth}
\put(-233,26){2$^{+}$}
\put(-233,73){4$^{+}$}
\put(-233,128){6$^{+}$}
\put(-233,185){8$^{+}$}
\put(-233,240){10$^{+}$}
\put(-208,280){{\bf TRVM}}
\put(-135,280){\bf exp.\cite{Kir}}
\put(-175,315){\large{\bf $^{130}$Ba g.s. band}}
\put(-65,280){{\bf CSM}}
\put(-225,315){{\large a)}}
\vfill\eject
\center
{\bf Figure 2b)}\\[24pt]
\epsfig{file=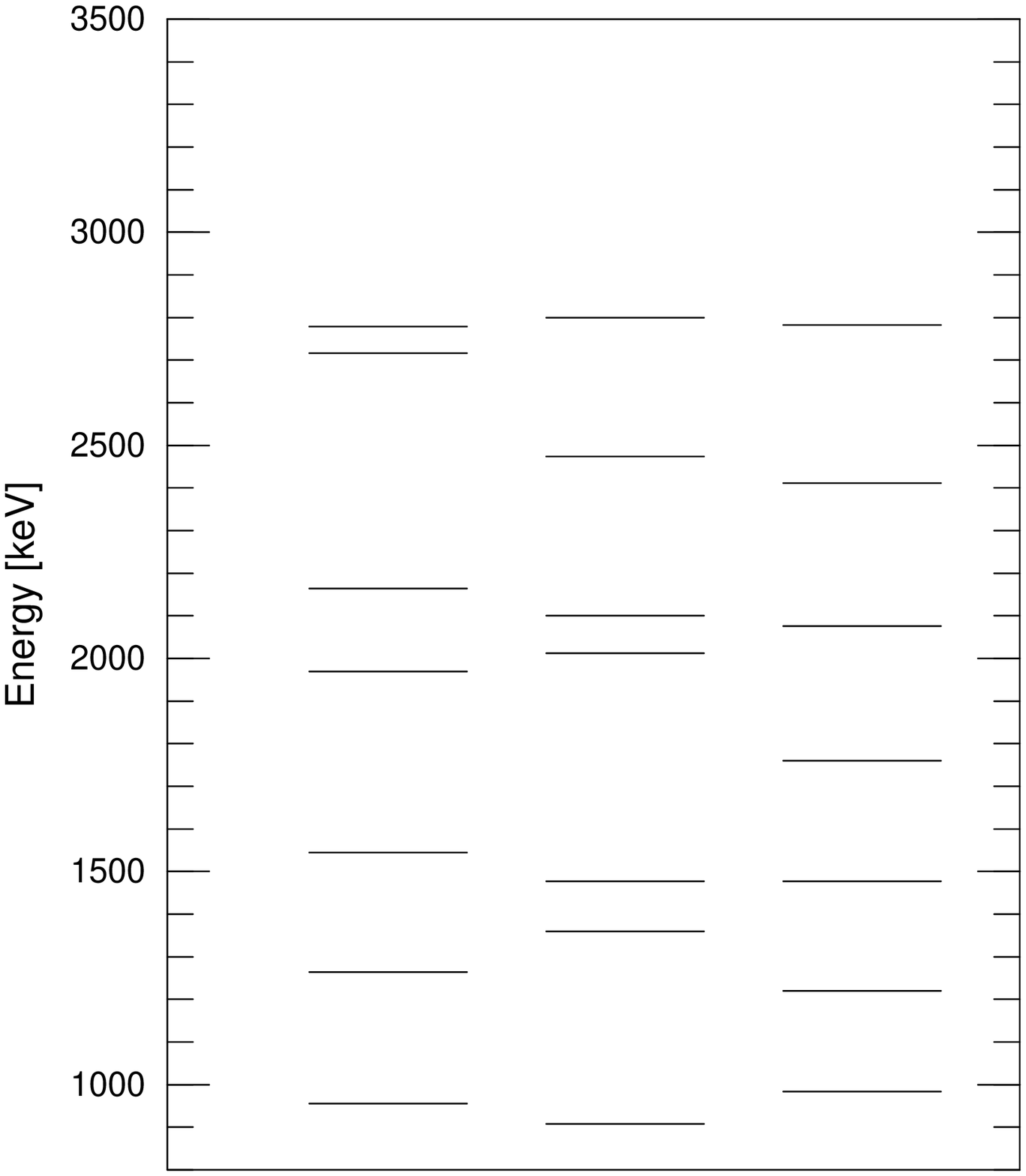,width=0.60\textwidth}
\put(-233,20){2$^{+}$}
\put(-233,55){3$^{+}$}
\put(-233,95){4$^{+}$}
\put(-233,135){5$^{+}$}
\put(-233,165){6$^{+}$}
\put(-233,220){7$^{+}$}
\put(-233,253){8$^{+}$}
\put(-208,280){{\bf TRVM}}
\put(-135,280){\bf exp.\cite{Kir}}
\put(-190,315){\large{\bf $^{130}$Ba $K$=2 $\gamma$ band}}
\put(-65,280){{\bf CSM}}
\put(-225,315){{\large b)}}
\vfill\eject
\center
{\bf Figure 3a)}\\[24pt]
\epsfig{file=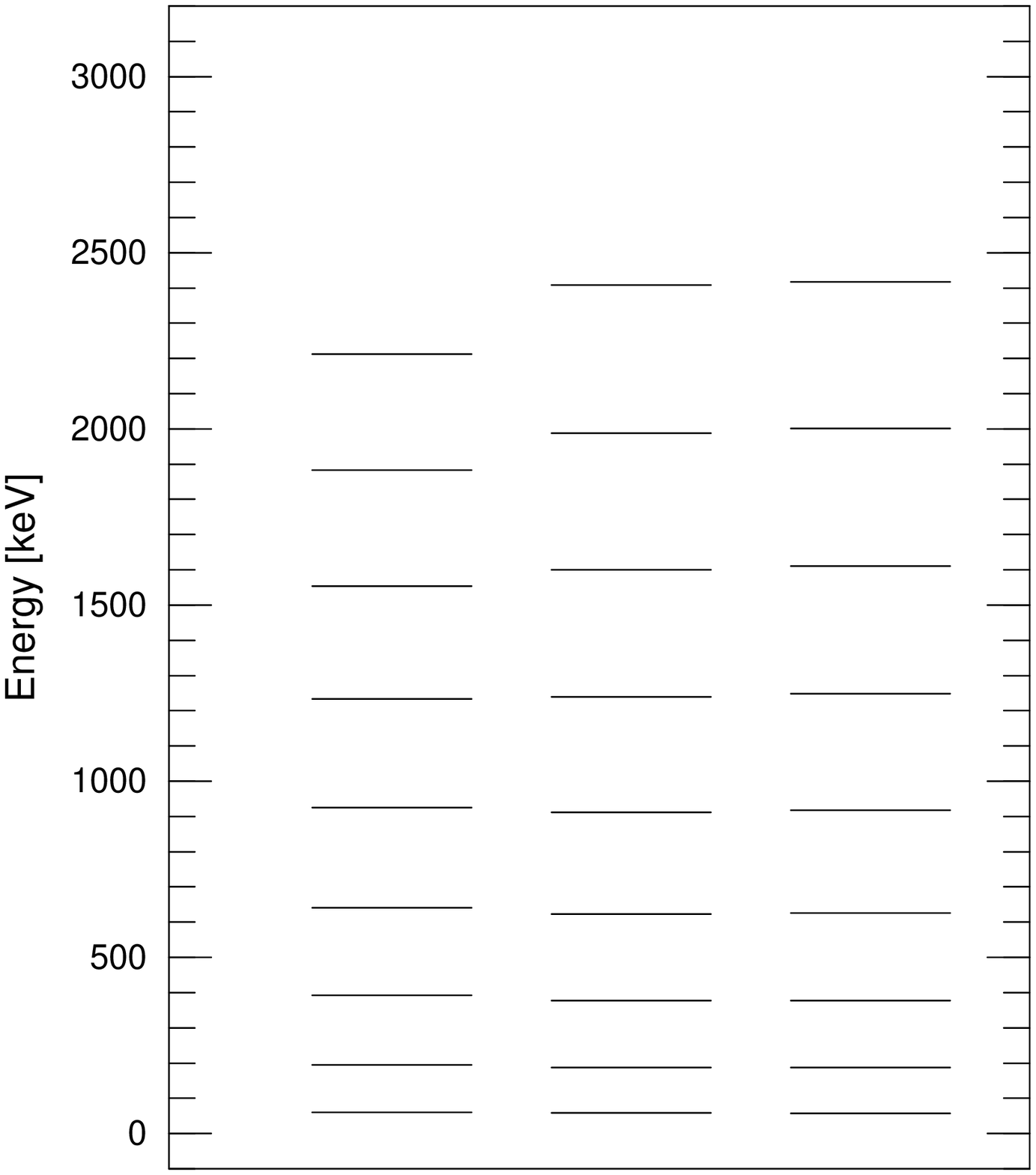,width=0.70\textwidth}
\put(-263,13){2$^{+}$}
\put(-263,35){4$^{+}$}
\put(-263,58){6$^{+}$}
\put(-263,87){8$^{+}$}
\put(-263,122){10$^{+}$}
\put(-263,160){12$^{+}$}
\put(-263,198){14$^{+}$}
\put(-263,243){16$^{+}$}
\put(-263,293){18$^{+}$}
\put(-242,325){{\large{\bf TRVM}}}
\put(-165,325){{\large{\bf exp.~\cite{gar98}}}}
\put(-75,325){{\large{\bf CSM}}}
\put(-200,358){{\large{\bf $^{228}$Th g.s. band}}}
\put(-260,360){{\large a)}}
\vfill\eject
\center
{\bf Figure 3b)}\\[24pt]
\epsfig{file=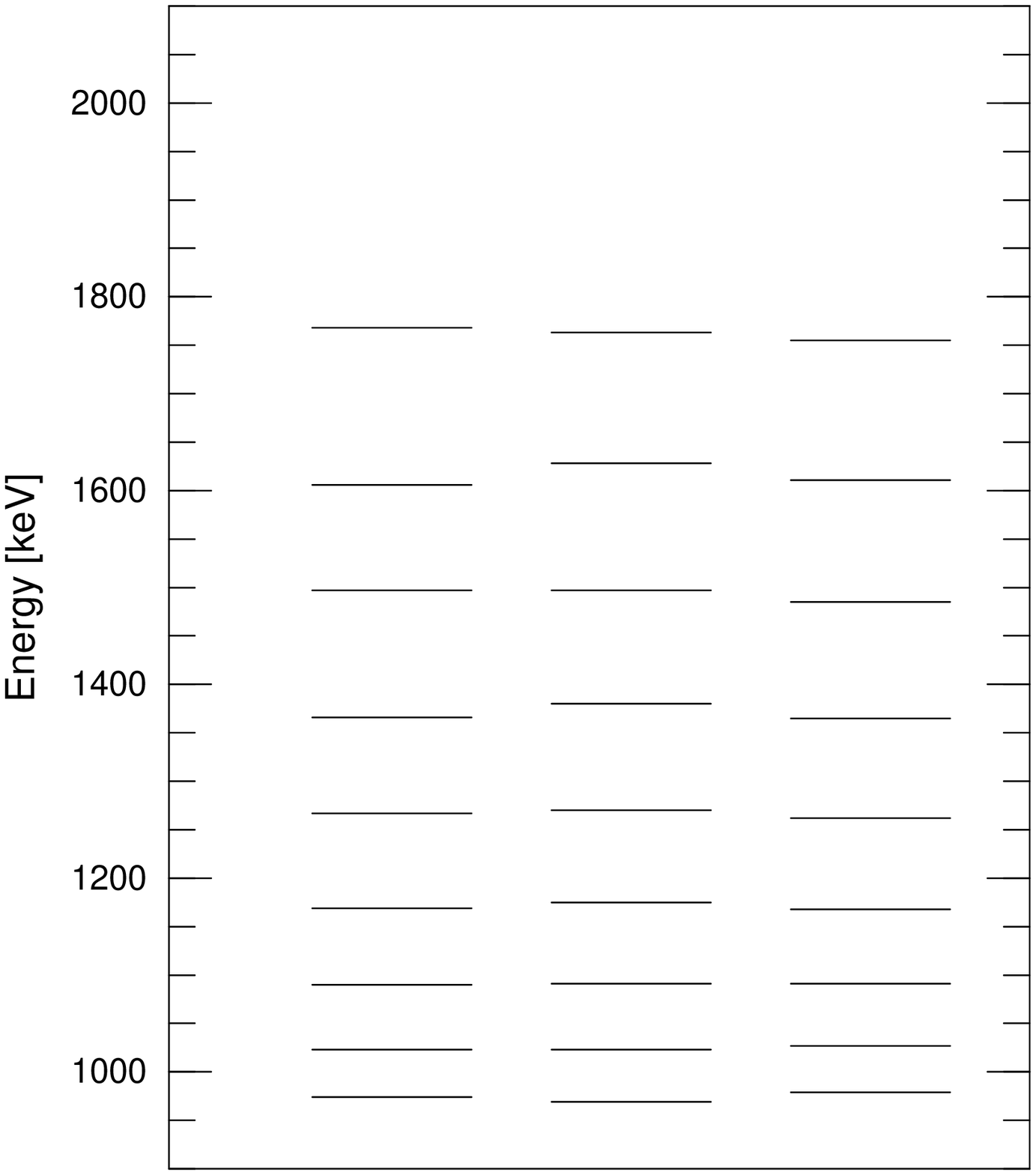,width=0.70\textwidth}
\put(-263,22){2$^{+}$}
\put(-263,42){3$^{+}$}
\put(-263,62){4$^{+}$}
\put(-263,88){5$^{+}$}
\put(-263,120){6$^{+}$}
\put(-263,153){7$^{+}$}
\put(-263,193){8$^{+}$}
\put(-263,235){9$^{+}$}
\put(-263,280){10$^{+}$}
\put(-242,325){{\large{\bf TRVM}}}
\put(-165,325){{\large{\bf exp.~\cite{gar98}}}}
\put(-205,358){{\large{\bf $^{228}$Th $K$=2 $\gamma$ band}}}
\put(-75,325){{\large{\bf CSM}}}
\put(-260,360){{\large b)}}
\end{document}